\documentclass[pra,twocolumn,showpacs,amsmath,amssymb,superscriptaddress]{revtex4}

\usepackage{color}
\usepackage{graphicx}
\usepackage{dcolumn}
\usepackage{bm}
\usepackage{hyperref}
\usepackage{mathrsfs}
\usepackage{soul}
\usepackage{ulem}
\usepackage{tabularx}
\usepackage{leftidx}
\usepackage[capitalize]{cleveref}
\usepackage{enumitem}

\setlist[itemize]{leftmargin=*}

\begin{document}


\title{Quantum key distribution over quantum repeaters with encoding: Using Error Detection as an Effective Post-Selection Tool}

\author{Yumang Jing}
\affiliation{School of Electronic and Electrical Engineering, University of Leeds, Leeds, LS2 9JT, U.K.}

\author{Daniel Alsina Leal}
\affiliation{School of Electronic and Electrical Engineering, University of Leeds, Leeds, LS2 9JT, U.K.}

\author{Mohsen Razavi}
\affiliation{School of Electronic and Electrical Engineering, University of Leeds, Leeds, LS2 9JT, U.K.}


\begin{abstract}
We propose a post-selection technique, based on quantum error detection, for quantum key distribution (QKD) systems that run over quantum repeaters with encoding. In such repeaters, quantum error correction techniques are used for entanglement distillation. By developing an analytical approach to study such quantum repeaters, we show that, in the context of QKD, it is often more efficient to use the error detection, rather than the error correction, capability of the underlying code to sift out cases where an error has been detected. We implement our technique for three-qubit repetition codes by modelling different sources of error in crucial components of the system. We then investigate in detail the impact of such imperfections on the secret key generation rate of the QKD system, and how one can use the information obtained during entanglement swapping and decoding stages to maximize the rate. For benchmarking purposes, we specify the maximum allowed error rates in different components of the setup below which positive key rates can be obtained.
\end{abstract}

      
\maketitle

\section{Introduction}
Quantum repeaters (QRs) are among the most complex systems that need to be developed for true quantum connectivity in the future \cite{kimble2008quantum,wehner2018quantum}. Their implementation faces experimental challenges in developing long coherent storage of quantum states and/or reliable processing of them, as well as efficient interfaces and measurement modules \cite{razavi2018introduction}. They would also give rise to theoretical conundrums of understanding and modelling multipartite entangled states in the presence of various imperfections. As a possible way forward, it would be interesting to come up with useful setups---implementable with today's or near future technologies---that benefit from concepts and techniques in QRs. Twin-field \cite{lucamarini2018overcoming,curty2019simple,zhong2019proof} and memory-assisted \cite{panayi2014memory,piparo2017memory,piparo2017measurement,bhaskar2020experimental} quantum key distribution (QKD) are among such examples. In this work, we look at an interesting class of quantum repeaters, which rely on quantum error correction for their entanglement distillation \cite{jiang2009}, and examine how best such systems can be used for QKD applications. We develop reliable tools to obtain and study the shared states between two users via such a repeater chain. This enables us to quantify how different components of the shared state contribute to the distillable secret key rate. This leads us to an interesting observation that majority of the obtained secret key bits correspond to the cases where no error has been detected along the repeater chain and decoder modules. This would then offer a simple, but effective, post-selection technique that considerably improves the key rate and increases the resilience of the system to errors as compared to when error correction is applied, as originally intended, with no post-selection.

Theoretically speaking, quantum repeaters have gone through a number of development stages, sometime referred to as different QR generations \cite{muralidharan2016optimal}. QRs were initially proposed to enable entanglement distribution, in an efficient way, at long distances \cite{briegel1998}. Using teleportation techniques~\cite{bennett1993teleporting,bouwmeester1997experimental}, one can then send quantum information across a quantum network once entangled states are shared between remote users. The main idea behind such repeaters is to split the link into shorter \textit{elementary} segments and first distribute and store entanglement over such links. One can then use entanglement swapping (ES)~\cite{zukowski1993event} and, possibly, entanglement distillation (ED) to establish high fidelity entanglement over long distances. In early implementations of QRs, it is expected that the initial entanglement distribution \cite{yu2020entanglement} over elementary links, as well as ES and ED operations to be probabilistic \cite{duan2001}. This can make the whole system too slow as many steps may need to be repeated upon a failure \cite{razavi2009quantum}. Such probabilistic QRs often require quantum memories with long coherence times, comparable to the transmission delay between the two end users \cite{piparo2013long}. It would help, to some extent, if the ES operation is deterministic, but, so long as the ED steps are probabilistic \cite{bennett1996purification,deutsch1996quantum}, we still need to repeat part of the protocol every time that ED fails. A remedy to this problem was proposed in  \cite{jiang2009}, in which ED is effectively done by using quantum error correction techniques. This operation can, in principle, be done deterministically, and, combined with deterministic ES operations, it can give a boost to the entanglement generation rate in a QR. This advantage, which corresponds to less waiting time, hence lower required coherence times, may come at the price of more demanding quantum processing requirements. By further improving our quantum error correction and quantum processing capabilities, one can design QR systems that are resilient to loss and error, hence can accomplish all the required tasks in a QR in a deterministic way \cite{muralidharan2014ultrafast, muralidharan2018one, glaudell2016serialized, munro2012quantum, fowler2010surface}. In such QRs, we no longer need to distribute entanglement. Instead, we can directly encode our message into a codeword and send it hop-by-hop across the network. This will, however, be further away in terms of an experimental demonstration. 

In this work, in the spirit of having an eye on near-future implementations, our focus will be on the transition from probabilistic QRs to deterministic QRs that use quantum error correction techniques only for their ED operations \cite{jiang2009,munro2010multiplexing,zwerger2014hybrid}. In such QRs, using a number of bipartite entangled states, we create a multi-qubit entangled codeword across elementary links. As we apply the ES operations, this codeword structure will then allow us to correct some of the errors that happen because of imperfections in the employed gates, measurement modules, and/or the initially distributed bipartite states. 

In principle, one can choose different code structures to implement such systems. Here, we choose the repetition codes to study and develop our methodology. They offer a simple structure, which can make their implementation easier, and still have relevance in systems where one type of error is more dominant than the other. For instance, if the memory decoherence is affected mainly by a dephasing process, the corresponding errors are modelled by the Pauli operator $Z$ \cite{panayi2014memory}, hence a code structure resilient to this type of error could be useful. The repetition codes would also offer a good learning platform, for theoretical studies, to better understand how different components of the system can affect the final result, and to come up with relevant techniques for analysing more complicated code structures. 

In this work, we devise an analytical method to study the above QR setups. We, however, realise that, even for the seemingly simple case of repetition codes, the analysis can become cumbersome quite quickly. Previous work on this subject \cite{jiang2009,bratzik2014} often rely on various approximations to analyse the system. In this work, we try to remain as close as we can to the exact results and only use approximations that are analytically justified and numerically verified. Our approach relies on the \textit{linearity} of the quantum circuits, and the \textit{transversality} of the code employed to manage the complexity of the analysis. This will enable us to obtain an accurate picture of the requirements of such systems in practice.

Using our methodology, we study the performance of QKD systems run over QRs with three-qubit repetition codes by accounting for various sources of error in the setup. We identify the terms that significantly impact the secret key generation rate, and then assess its dependence on relevant error parameters. In previous work on this subject \cite{bratzik2014}, the repeater chain is used to create a bipartite entangled state, which the two users will then employ to exchange a secret key. In our work, we allow the users to exploit the information obtained during the entanglement swapping and decoding stages to divide the states that they obtain, and keys generated from them, into different groups. This not only improves the key rate and the resilience of the system to errors, but also allows us to identify states that majorly contribute to the secret key rate. It turns out that in most cases the key contribution is from the \textit{golden} states for which no error has been detected at either swapping or decoding stage. This will enable us to use an efficient post-selection technique that not only simplifies the analysis of the system, but also can reduce the complexity in any practical demonstration of the setup. We believe that our work can pave the way for similarly detailed analysis of other repeater protocols with more complex encoding. This will enable quantitative rate-versus-resource analysis for various protocols.

The paper is organized as follows. In Sec.~\ref{schematic_section}, we begin with a description of the repeater protocol in Ref.~\cite{jiang2009} and the error models we use to formulate the problem in hand. In Sec.~\ref{first_nesting_section}, we discuss the {linearization} method employed for our study and go over the exact analysis for nesting level one. We fully study the effect of different terms, components, and system imperfections before generalizing our results, in Sec.~\ref{higher_nesting_level_section}, to higher nesting levels. We present the dependence of the secret key generation rate for such QRs on different error parameters and find the corresponding thresholds for extracting a nonzero secret key rate at different nesting levels. Finally, we conclude the paper in Sec.~\ref{conclusion_section}.

\section{System Description}
\label{schematic_section}

In this section, we first start with a detailed review of the QR scheme with encoding proposed by Jiang \textit{et al.} \cite{jiang2009} and the respective quantum circuits designed to implement it. Then, we introduce the error models considered in our analysis, followed by the problem statement and the key objectives of our study. 

In this work, we mainly use the 3-qubit repetition code as an example to illustrate and develop our key ideas and techniques, where the logical qubits are encoded as
\begin{align}
| \tilde{0} \rangle = |000\rangle \quad \text{and} \quad |\tilde{1}\rangle = |111\rangle,
\label{logical01}
\end{align}
where $|0\rangle$ and $|1\rangle$ represent the standard basis for a single qubit. This code can correct up to one bit-flip error. Although it is not a strong error correction code, the thorough analysis of its performance with possible errors considered in its implementation will still offer us an indication of how this type of QRs performs. 
\subsection{Quantum repeater with 3-qubit repetition code}
\label{Sec:3quRep}
\begin{figure}[t]
\includegraphics[width=\columnwidth]{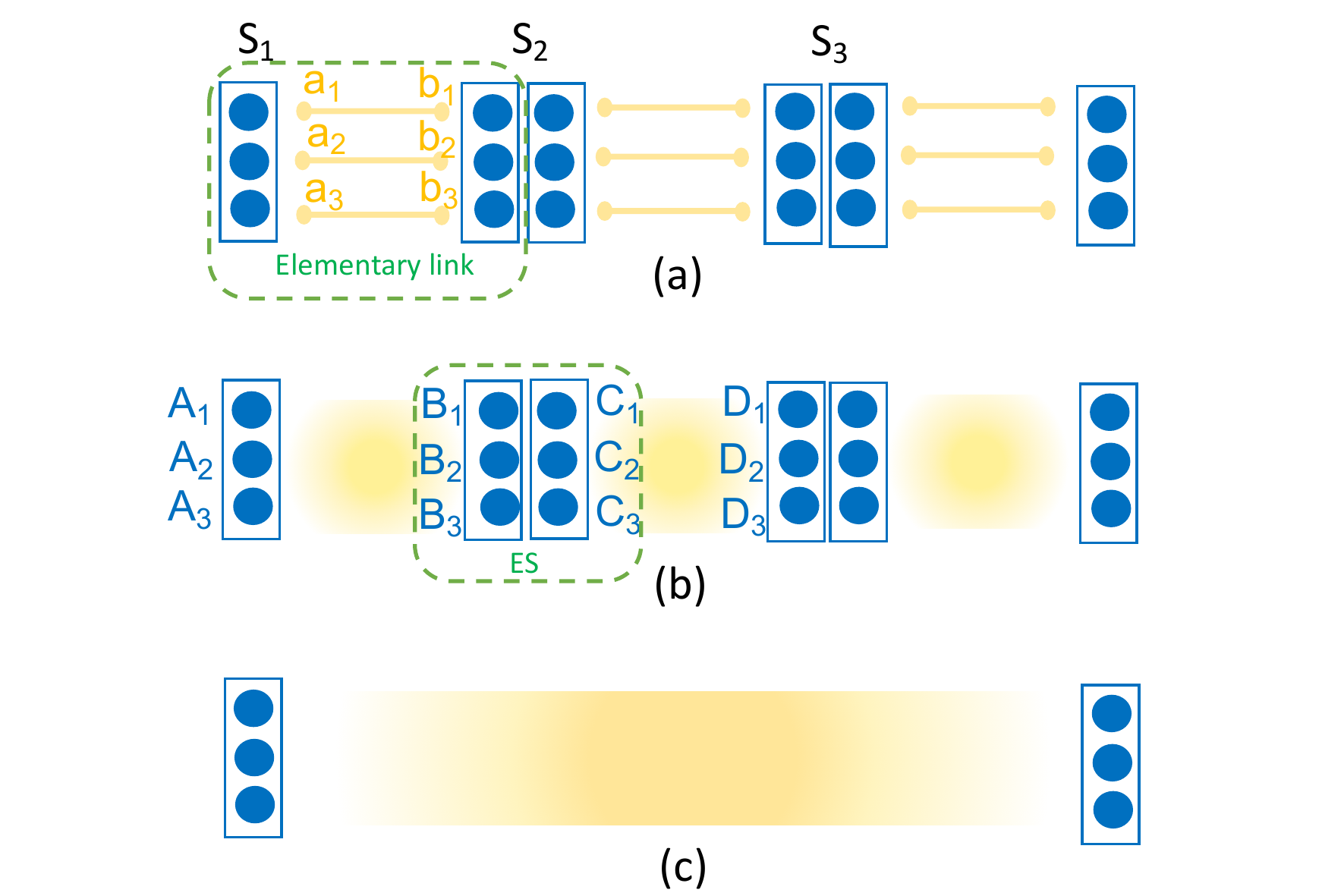}
\caption{\label{schematic} Schematic representation of quantum repeaters with encoding. (a) The codeword states are locally prepared at each memory bank (large blue ovals) and original Bell pairs are distributed between neighboring nodes (small yellow ovals connected by yellow lines); (b) Encoded Bell pairs are generated between neighbouring stations by performing remote CNOT gates; (c) The encoded ES operations are performed at each intermediate station simultaneously. This creates an encoded Bell pair between the two end users. Based on the measurement results at each middle node, the Pauli frame of the final entangled state can be adjusted.}
\end{figure}
Here, we describe the ideal setting of the protocol proposed in \cite{jiang2009} in the special case of 3-qubit repetition codes. In this protocol, depicted schematically in Fig.~\ref{schematic}, we first generate encoded entangled states across \textit{all} elementary links, and then apply ES operations at intermediate nodes to both distill and swap entanglement across the chain. 

The encoded entangled state of interest across an example elementary link $A$-$B$ in Fig.~\ref{schematic} is in the form 
\begin{equation}
| \tilde{\Phi}^{+} \rangle_{A,B}=\frac{1}{\sqrt{2}}(|\tilde{0}\rangle_A |\tilde{0}\rangle_B+| \tilde{1}\rangle_A | \tilde{1}\rangle_B),
\end{equation}
where $|\tilde{i}\rangle_K \equiv |i\rangle_{K_1}|i\rangle_{K_2}|i\rangle_{K_3}$, for $i=0,1$ and $K=A,B$. In Fig.~\ref{schematic}, the memory bank $K = \{K_1,K_2,K_3\}$ is shown by large (blue) ovals. This multipartite entangled state is distributed between memory banks $A$ and $B$ in the following way:

\textit{Step 1} Initialize memory banks $A$ and $B$ in the codeword states $\frac{1}{\sqrt{2}} (| \tilde{0} \rangle_A + | \tilde{1} \rangle_A)$ and $|\tilde{0}\rangle_B$, respectively. The codeword state for node $A$ can be achieved by applying two controlled NOT (CNOT) gates, $\text{CNOT}_{A_1 \rightarrow A_2}$ and $\text{CNOT}_{A_1 \rightarrow A_3}$, on the state $\frac{1}{\sqrt{2}} (|0\rangle_{A_1} + |1\rangle_{A_1}) |0\rangle_{A_2} |0\rangle_{A_3}$, where, in the notation $\text{CNOT}_{K \rightarrow J}$, $K$ is the control qubit and $J$ is the target qubit. We use the same notation for pairwise CNOT gates between qubits in two memory banks $K$ and $J$. This ideally leads to the desired codeword state
\begin{equation}
\frac{1}{\sqrt{2}} (|000\rangle_A + |111\rangle_A) = \frac{1}{\sqrt{2}} (| \tilde{0} \rangle_A + | \tilde{1} \rangle_A). 
\label{perfectencoding}
\end{equation}
The above state can, in principle, be obtained probabilistically as well, by repeating a preparation procedure until success. Given that the preparation is a local process, it can possibly be repeated at a sufficiently fast rate to ensure success in a  reasonable time.

\textit{Step 2} In order to generate $| \tilde{\Phi}^{+} \rangle_{A,B}$, we share 3 bipartite maximally entangled states between the corresponding memories in memory banks $a$ and $b$, shown by small yellow ovals in Fig.~\ref{schematic}(a), co-located, respectively, with memory banks $A$ and $B$. These Bell states, shown by yellow lines in Fig.~\ref{schematic}(a), can be distributed in advance, or in parallel with step 1. The implementation of this process and the quality of the generated entangled states depend on the specifics of the employed experimental platform. Normally, this step is mediated with photons, hence is often probabilistic and needs to be heralding. 

\begin{figure}[t]
\includegraphics[width=.8\columnwidth]{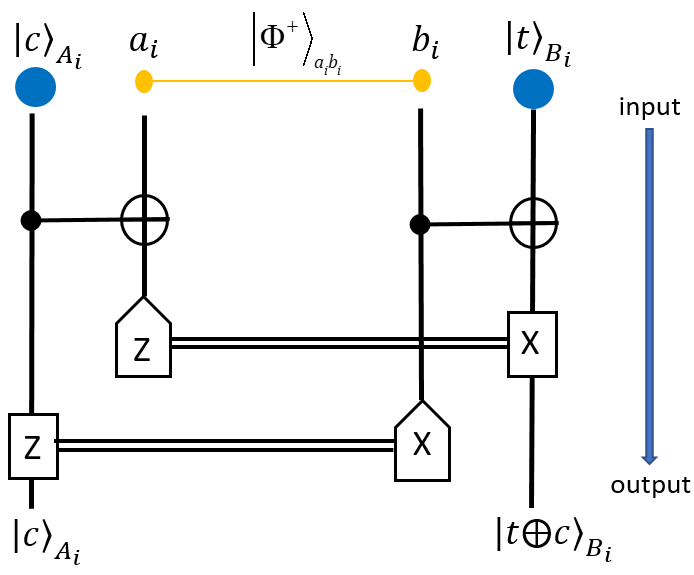}
\caption{\label{cnot}  Circuit for remote CNOT gate \cite{jiang2007distributed} between a qubit $c$ at node $A_i$, as control qubit, and a quibt $t$ at node $B_i$, as the target qubit. Using the maximally entangled state shared between $a_i$ and $b_i$ nodes, and by applying two local CNOT gates on $A_i$-$a_i$ and $b_i$-$B_i$ pairs, we can effectively implement a remote CNOT gate on $A_i$-$B_i$. Note that this requires single-qubit measurements on $a_i$ and $b_i$, classical communication, and local single-qubit rotation on $A_i$ and $B_i$.}
\end{figure}

\textit{Step 3} We use the distributed bipartite entangled states to implement three remote CNOT gates, see Fig.~\ref{cnot} and its caption for further detail, which are applied transversally, leading to the desired state for the elementary link: 
\begin{align}
    \frac{1}{\sqrt{2} }(|\tilde{0}\rangle_A + |\tilde{1}\rangle_A)\otimes |\tilde{0}\rangle_B \stackrel{\text{CNOT}_{A \rightarrow B}}{\longrightarrow} | \tilde{\Phi}^{+} \rangle_{A,B}.
\end{align}

Once the encoded entangled states are distributed across all elementary links, the next step is to perform ES operations at all intermediate stations to extend the entanglement to the entire link. For instance, in Fig.~\ref{schematic}(b), in order to establish multipartite entanglement between memory banks $A$ and $D$, we perform an encoded Bell measurement on memory banks $B$ and $C$. This, due to the transversality of the employed code, is simply done by performing three individual Bell-state measurements (BSMs) on the corresponding pairs of physical qubits in $B$ and $C$. More specifically, such BSMs can be realized deterministically by applying $\text{CNOT}_{B_i \rightarrow C_i}$, followed by a projective X-measurement on qubit $B_i$ and Z-measurement on qubit $C_i$, for $i=1,2,3$. In the ideal case, right before the single-qubit measurements, the initial state of the two links would then undergo the following transformation \cite{jiang2009}
\begin{align}
&| \tilde{\Phi}^{+} \rangle_{A,B} \otimes | \tilde{\Phi}^{+} \rangle_{C,D} \nonumber\\
= {} & \frac{1}{2}( |\tilde{\Phi}^{+} \rangle_{A,D} \otimes | \tilde{\Phi}^{+} \rangle_{B,C} + |\tilde{\Phi}^{-} \rangle_{A,D} \otimes | \tilde{\Phi}^{-} \rangle_{B,C} \nonumber\\
                     & +|\tilde{\Psi}^{+} \rangle_{A,D} \otimes | \tilde{\Psi}^{+} \rangle_{B,C} + |\tilde{\Psi}^{-} \rangle_{A,D} \otimes | \tilde{\Psi}^{-} \rangle_{B,C}) \nonumber\\
\stackrel{\text{CNOT}_{B \rightarrow C}}{\longrightarrow} & \frac{1}{2}( |\tilde{\Phi}^{+} \rangle_{A,D} \otimes | \tilde{+} \rangle_{B} | \tilde{0} \rangle_{C} + |\tilde{\Phi}^{-} \rangle_{A,D} \otimes | \tilde{-} \rangle_{B} | \tilde{0} \rangle_{C} \nonumber\\
                     & +|\tilde{\Psi}^{+} \rangle_{A,D} \otimes | \tilde{+} \rangle_{B} | \tilde{1} \rangle_{C}+ |\tilde{\Psi}^{-} \rangle_{A,D} \otimes | \tilde{-} \rangle_{B} | \tilde{1} \rangle_{C}),  
                \label{eq_es}       
        \end{align}
where $| \tilde{\Phi}^{\pm} \rangle_{A,D}= \frac{1}{\sqrt{2}} ( |\tilde{0} \rangle_{A} |\tilde{0} \rangle_{D} \pm  |\tilde{1} \rangle_{A} |\tilde{1} \rangle_{D})$, $| \tilde{\Psi}^{\pm} \rangle_{A,D}= \frac{1}{\sqrt{2}} ( |\tilde{0} \rangle_{A} |\tilde{1} \rangle_{D} \pm  |\tilde{1} \rangle_{A} |\tilde{0} \rangle_{D})$ and 
\begin{align}
 |\tilde{\pm}\rangle_{B}&= \frac{1}{\sqrt{2}}(|\tilde{0} \rangle_{B} \pm |\tilde{1} \rangle_{B}) \nonumber \\
 &= \frac{1}{2} (|\pm\pm\pm\rangle_B + |\pm\mp\mp\rangle_B  \nonumber \\
 &\, \,\, + |\mp\pm\mp\rangle_B + |\mp\mp\pm\rangle_B),
\end{align} 
with $|\pm\rangle = (|0\rangle \pm |1\rangle)/\sqrt{2}$ for a single qubit.

By measuring the states of $B_i$ and $C_i$, $i=1,2,3$, in, respectively, X and Z basis, we project the state of $A$ and $D$ in \cref{eq_es} into one of the encoded Bell states. This becomes possible because all terms in $|\tilde{+} \rangle$ ($|\tilde{-} \rangle$) have an odd (even) number of $|+\rangle$ states, and measuring $|\tilde{0}\rangle$ ($|\tilde{1}\rangle$) ideally results in three $|0\rangle$ ($|1\rangle$) states. In the non-ideal case, it is possible that, instead of three identical outputs, we get, for instance, two $|0\rangle$s and one $|1\rangle$. But, then, because of the employed error correction scheme, we can still identify which Bell state is the most likely outcome of the ES process. Note that, by accounting for the erroneous cases, there will be $64$ different combinations of measurement outcomes, and each of them will uniquely lead to one type of encoded Bell pair. Even though the measurement outcomes at each middle station should be announced to Alice and Bob to determine the Pauli frame for the encoded Bell pair shared by them in the end, this scheme would not rely on any communication among middle stations, which reduces the time scaling from polynomial to polylogrithmic \cite{muralidharan2016optimal}. For applications, such as QKD, that can deal with Pauli frame adjustments at the post-measurement stage, this scheme also lowers the waiting time, and correspondingly the required coherence time for the memories.

\begin{figure}[t]
\includegraphics[width=.8\columnwidth]{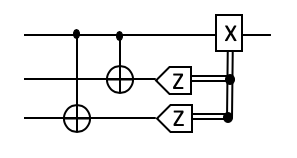}
\caption{\label{decoding}  Quantum circuit for decoding 3-qubit repetition codes \cite{bratzik2014}. Alice and Bob will both use the same circuit for decoding, in which they flip their first qubit if they measure $|1\rangle$ in all other qubits. }
\end{figure}

After all ES operations, an encoded entanglement is ideally distributed between the two end users. In order to do QKD, or other possible applications, the final encoded entangled states can be decoded into a bipartite state. The decoding circuit employed in this work is simply the reverse process of the encoding procedure for three-qubit repetition codes \cite{bratzik2014}, as depicted in Fig.~\ref{decoding}. Alice and Bob each apply this circuit to their three qubits in hand, and measure two of them. They flip the first qubit only if they measure $|1\rangle$ in the other two qubits. 

The above repeater protocol implements an implicit entanglement distillation by using error correction techniques. This is partly done at the ES stage and is supplemented by the final error correction that happens at the decoding stage. But, for protocols such as QKD, which can cope with discarding data if needed, the other possibility is to use the information available at the ES stations to discount the end-to-end distributed state if an error has been detected at any intermediate stage. By doing so, we only keep the cases for which we are more confident that we have got the desired Bell state, and, effectively, distill the entanglement generated by the repeater chain. So long as the chance of error is low, this still offers a nearly deterministic solution for quantum repeaters. 

In this work, we will examine how the above idea can improve the performance of QKD systems that run over such repeaters. It turns out that the secret key rate of such QKD systems is dominated by the post-measurement state corresponding to when no error has been detected at ES and decoding stages. Nevertheless, we still need to calculate the effect of errors on system performance. Detecting no errors by our error correction scheme does not guarantee the absence of errors. The decoded state is still affected by errors not detectable by our error correction scheme, some of which would correspond to higher order error terms that may not be properly accounted for if our analysis is not sufficiently accurate. In the following, we first summarise the error models used in our analysis. We then describe the problem in the context of previous research on this subject. 

\subsection{Error models}
\label{error_model_section}
Three major imperfections are considered in our analysis:\\ 
(1) \textbf{Imperfections in initial Bell states:} The originally distributed Bell states, i.e., yellow links in Fig.~\ref{schematic}(a), are not necessarily perfect. We model them as Werner states with fidelity $F_0$:
\begin{align}
\rho^{\rm W}=F_0 | \phi^{+} \rangle \langle \phi^{+} | + \frac{1-F_0}{3} (\mathbb{I}_4-| \phi^{+} \rangle \langle \phi^{+} |),
\label{original_bell}
\end{align}
where $| \phi^{+} \rangle= \frac{1}{\sqrt{2}} ( |00\rangle + |11\rangle )$ and $\mathbb{I}_4$ is a $4 \times 4$ identity matrix.\\
(2) \textbf{Gate imperfections:} We employ the generic model for imperfect two-qubit operations introduced in Ref.~\cite{briegel1998}. The unitary operation $U_{i,j}$, acting on qubits $i$ and $j$, is modelled by 
\begin{align}
\label{eq:CNOTmodel}
\rho^{\text{out}} = (1-\beta) U_{i,j} \rho^{\text{in}} U_{i,j}^{\dagger} + \frac{\beta}{4} \text{Tr}_{i,j} (\rho^{\text{in}}) \otimes \mathbb{I}_{i,j},
\end{align} 
where $\rho^{\text{in}}$ ($\rho^{\text{out}}$) is the input (output) before (after) the two-qubit gate $U_{i,j}$, $\beta$ is the gate error probability and $\mathbb{I}_{i,j}$ is the identity matrix for qubits $i,j$. The main two-qubit gate used in this paper is $\text{CNOT}_{i \rightarrow j}$. \\
(3) \textbf{Measurement imperfections:} The projective measurements to states $| 0\rangle$ and $| 1\rangle$ are, respectively, represented by 
\begin{align}
\label{eq:meas}
P_0 &= (1-\delta) |0\rangle \langle 0| + \delta |1\rangle \langle 1| \quad \mbox{and} \nonumber\\
P_1 &= (1-\delta) |1\rangle \langle 1| + \delta |0\rangle \langle 0|,
\end{align}
where $\delta$ is the measurement error probability. Similar measurement operators, $P_\pm$, are used for projective measurement in $| \pm \rangle$ basis. 

In our analysis, we neglect other types of errors that may be present in a real setup. For instance, we assume all single-qubit unitary operations, i.e., bit-flip (X gate) or phase-flip (Z gate) rotations, are perfect. In the case of QKD as an application, this is justified as these operations can typically be implemented in the classical post-processing stage. In order to simplify the analysis, we also assume that quantum memories with sufficiently long coherence times are available. Considering that the waiting time for encoded QRs is comparatively low, we neglect the memory decoherence effects in our analysis. 

\subsection{Problem Description}
In this work, we study the performance of a QKD system that is run over an encoded QR setup with three-qubit repetition codes by accounting for errors in the setup as presented above. We consider an entanglement-based QKD setup that relies on BBM92 protocol \cite{bennett1992quantum}. We use an asymmetric implementation of the protocol where the two end users, Alice and Bob, choose the two measurement bases, i.e., Z and X bases, unevenly, in order to increase the basis-sift factor \cite{lo2005efficient}. Our objective is to assess the dependence of the secret key generation rate in our QKD system on relevant error parameters. To this end, we first need to calculate the secret key generation rate per decoded state, $\rho^{\rm dec}$, shared between Alice and Bob. In the asymptotic regime, this parameter, known as secret fraction \cite{bratzik2014}, is given by \cite{Shor-PreskillBB84}
\begin{equation}
    r_{\infty}(\rho^{\rm dec})=\text{max}\{ 0,1-h(e_z)-h(e_x) \},
\label{secfraction}
\end{equation}
where $h(p)=-p \text{log}_2 p - (1-p) \text{log}_2 (1-p)$ is the binary Shannon entropy, and
\begin{align}
    e_z &= {\rm Tr}(P_0^{\rm Alice}P_1^{\rm Bob}\rho^{\rm dec}) + {\rm Tr}(P_1^{\rm Alice}P_0^{\rm Bob}\rho^{\rm dec})\nonumber \\
    &=  (\delta^2 + (1-\delta)^2) (\langle\psi^+|\rho^{\rm dec}|\psi^+\rangle + \langle\psi^-|\rho^{\rm dec}|\psi^-\rangle) \nonumber \\
    &\, +  2\delta (1-\delta)(\langle\phi^+|\rho^{\rm dec}|\phi^+\rangle + \langle\phi^-|\rho^{\rm dec}|\phi^-\rangle) \nonumber \\
     e_x &= {\rm Tr}(P_+^{\rm Alice}P_-^{\rm Bob}\rho^{\rm dec}) + {\rm Tr}(P_-^{\rm Alice}P_+^{\rm Bob}\rho^{\rm dec})\nonumber \\
     &= (\delta^2 + (1-\delta)^2)(\langle\phi^-|\rho^{\rm dec}|\phi^-\rangle + \langle\psi^-|\rho^{\rm dec}|\psi^-\rangle) \nonumber \\
     &\, + 2\delta(1-\delta) (\langle\phi^+|\rho^{\rm dec}|\phi^+\rangle + \langle\psi^+|\rho^{\rm dec}|\psi^+\rangle)
\label{eq:ezex}
\end{align}
are, respectively, the observed error rates in Z and X bases, where $|\phi^{\pm} \rangle= \frac{1}{\sqrt{2}} ( |00\rangle \pm |11\rangle )$ and $| \psi^{\pm} \rangle= \frac{1}{\sqrt{2}} ( |01\rangle \pm |10\rangle )$, in the joint state space of Alice and Bob, and measurement operators are defined according to \cref{eq:meas} with additional superscripts to specify the affected qubit.

In order to understand the effect of various system parameters on the final secret key rate, we simulate the above setting in the nominal mode of operation where no eavesdropper is present. In this case, $\rho^{\rm dec}$ will then be given by the shared state between Alice and Bob after decoding, from which we can calculate the error parameters $e_z$ and $e_x$ in the asymptotic regime, where an infinite number of entangled states are shared among users. Our problem would then reduce to specifying what $\rho^{\rm dec}$ is in a typical error-prone QR setting with encoding.

While at first glance this may look like a quite straightforward problem, in practice, we face some computational challenges. The obvious way to calculate the final entangled state is to obtain the encoded entangled state at each elementary link and then apply ES in a nested way. For a 3-qubit repetition code, the ES operation involves 12 qubits, so our operation is on a space with dimension $2^{12}$. This may sound manageable, but certainly not scalable. The next simplest code, i.e., 5-qubit repetition code, requires operation on 20 qubits, or a space of dimension $2^{20}$. It is easy to see how problem can get out of hand quite quickly. Proper analytical and numerical techniques are then needed to handle this problem.

Previous work on this subject \cite{jiang2009,bratzik2014} often rely on various approximations to solve the problem. The original work in \cite{jiang2009} makes some assumptions on how the initial states are prepared, based on which they estimate how much error, to the first order, is expected in each qubit. They then use their method to approximate the fidelity of the final state. While a good approach to prove the scaling improvement offered by their proposed scheme, it falls short of the accurate scheme that we need for key rate calculations. A follow-up paper by Bratzik \textit{et al.} \cite{bratzik2014} attempted to fill this gap by approximating the actual state that one would obtain for the decoded state of a 3-qubit repetition code by accounting for imperfections in the CNOT gates as well as the initial Bell states. They use several approximations to achieve this goal:
\begin{itemize}
    \item They model the error in a cascade of operations by separating the ideal and the first-order error term in the output from the rest, where the rest is modelled by a generic identity operator at the output. The first-order error term is modelled by the identity operator for the involved qubits in the operation.
    \item They find a set of operations that will be corrected by the BSM operation, in addition to what may be corrected by the employed code. Based on this, they find a set of correctable states that will be mapped to the desired encoded Bell states. They use these states to crudely calculate the probability of obtaining the desired state after a number of ES operations, and assume that, in all other cases, the identity operator is obtained. 
\end{itemize}
Based on the above assumptions, they would then conclude that the considered encoded QR cannot beat the original QR protocol in \cite{briegel1998} in terms of the achievable key rate or the required gate error parameters.

In this work, we improve upon the approach taken in \cite{bratzik2014} in several respects. First, we improve the accuracy of the calculations by accounting for errors in each gate individually rather than modelling the overall effect, for a cascade of gates, in a crude way. Our new approach enables us to show that the encoded QRs are resilient to larger margins of error than previously thought. It is also easier to apply our method to other codes than the 3-qubit repetition code considered in \cite{bratzik2014}, as some of their steps are specific to this employed code. As such, extending their approach to other code structures is not necessarily straightforward. Here, we employ an analytical approach that relies on the \textit{linearity} of the quantum circuits and the \textit{transversality} of the employed code, and, in principle, can be applied to other moderate code structures. Finally, an important element of our key rate analysis is to use the information reported by the middle nodes of the repeater chain at its end nodes. This allows us to classify the decoded states, based on the measurement results at the ES and decoding stages, resulting in a considerable improvement of system performance. This also reduces the complexity of the corresponding key rate analysis. Overall, this work enables us to obtain a more accurate picture of the requirements of such systems in practice, and whether, any simplified version of them, can realistically be built with current technologies.

In the following sections, we will first use the simplest repeater setup, where only one swap operation is performed, to describe our methodology, and to justify certain simplifying assumptions that we make in neglecting the less dominant terms. Then, we will extend our results to higher nesting levels and obtain the secret key rate in our setup as a function of various system parameters. 

\section{Methodology and Performance: Nesting level one}
\label{first_nesting_section}

In this section, we look at the simplest repeater setup with only one middle node corresponding to nesting level one. The initial objective here is to find a scalable methodology by which the final entangled state shared by Alice and Bob can be calculated. We then find the secret fraction corresponding to different decoded states conditioned on the measurement results at the ES and decoding stages. This allows us to better understand how each term and each imperfection affect system performance. This guides us toward finding simple, but still tight, approximations that reduce the complexity of the problem in hand.

\subsection{Linearization}
\label{linearization_section}
Our first objective is to develop a methodology to calculate the joint state between memory banks $A$ and $D$, $\rho_{AD}$, in \cref{schematic}(b), after one round of entanglement swapping. We first explain this procedure when the initial codeword states in memory banks $A$--$D$ are perfectly encoded as follows
\begin{align}
    \rho^{\text{in}}_{A} &=\frac{1}{2}(|000\rangle_A + |111\rangle_A)({}_{A}{\langle 000|} + {}_{A}{\langle 111|}) \nonumber \\
              &= \frac{1}{2} [ ( |0\rangle_A \langle 0| )^{\otimes 3} + ( |0\rangle_A \langle 1| )^{\otimes 3} \nonumber \\
              &+( |1\rangle_A \langle 0| )^{\otimes 3} + ( |1\rangle_A \langle 1|)^{\otimes 3}], \nonumber\\
    \rho^{\text{in}}_{B} &=|000\rangle_B \langle 000|= (|0\rangle_B \langle 0|)^{\otimes 3},
\label{linear}
\end{align}
where $(|i\rangle_K \langle j|)^{\otimes 3} \equiv |i\rangle_{K_1} \langle j| \otimes |i\rangle_{K_2} \langle j| \otimes |i\rangle_{K_3} \langle j|$, for $i,j=0,1$. The initial state for $C$ and $D$, $\rho^{\text{in}}_{C}$ and $\rho^{\text{in}}_{D}$, are, respectively, similar to that of $A$ and $B$. In this case, we can first find the joint state $\rho_{AB}$ ($\rho_{CD}$) of memory banks $A$ and $B$ ($C$ and $D$) after the remote CNOT operation, and then apply the ES operation. In this case, we have 
\begin{align}
\label{eq:rhoAB}
     \rho_{AB}^r= \frac{U_{AB}^r \text{Tr}_{ab} [M^r_{\rm RC} \mathcal{E}_{\rm RC}(\rho^{\rm in}) ] U_{AB}^r}{ \text{Tr} [M^r_{\rm RC} \mathcal{E}_{\rm RC}(\rho^{\rm in}) ]},
 \end{align}
 where $\text{Tr}_{ab}$ is the partial trace over memory banks $a$ and $b$, $\mathcal{E}_{\rm RC}$ is the combination of all remote CNOT gate operations on $Aa$ and $bB$ memory banks, $M^r_{\rm RC}$ is the collective projective measurement operator at this step corresponding to the pattern of measurement results given by $r$, and $U_{AB}^r$ is the corresponding Pauli frame correction in \cref{cnot}. In \cref{eq:rhoAB}, the input state is given by
\begin{equation}
\rho^{\rm in} = \rho^{\rm in}_A \otimes \rho^{\rm in}_B \otimes \rho^{\rm W}_{ab},        
\end{equation}
where $\rho^{\rm W}_{ab} = \rho^{\rm W}_{a_1b_1} \otimes \rho^{\rm W}_{a_2b_2} \otimes \rho^{\rm W}_{a_3b_3}$ as given by \cref{original_bell} for the subsystems specified by the subscripts. The quantum operation $\mathcal{E}_{\rm RC}$ is also given by
\begin{equation}
\label{eq:E}
    \mathcal{E}_{\rm RC} = \mathcal{E}_1 \otimes \mathcal{E}_2 \otimes \mathcal{E}_3,
\end{equation}
with, for $i=1,2,3$,
\begin{equation}
\label{eq:Ei}
    \mathcal{E}_i = \mathcal{E}_{A_ia_i} \otimes \mathcal{E}_{b_iB_i} , 
\end{equation}
where $\mathcal{E}_{KJ}$ is given by the transformation in \cref{eq:CNOTmodel} for the gate $\text{CNOT}_{K \rightarrow J}$. 

As mentioned earlier the direct approach of calculating $\mathcal{E}_{\rm RC}(\rho^{\rm in})$ requires dealing with a space of dimension $2^{12}$ even for the simple 3-qubit repetition code considered here. In order to simplify the process and reduce the time required for running the code, we use the linearity of operator $\mathcal{E}_{\rm RC}$ and its tensor product form in \cref{eq:E}. To be more precise, using \cref{linear}, we have
\begin{align}
\label{ERC-sum}
    \mathcal{E}_{\rm RC}(\rho^{\rm in}) = \frac{1}{2} \sum_{j,k=0,1} {\bigotimes_{i=1}^3{\mathcal{E}_i(|j\rangle_{A_i}\langle k|\otimes |0\rangle_{B_i}\langle 0|\otimes \rho^{\rm W}_{a_ib_i})}} .
\end{align}
By the above trick, we reduce the computational complexity of the problem to effectively that of a 4 qubit system in each row comprising of qubits $A_i$, $a_i$, $b_i$, and $B_i$, for $i=1,2,3$. For each component of the input state, we just need to calculate the output for one row, extend it to all rows by a simple tensor product, and then sum over all possible input components. 

In order to calculate $\rho_{AB}^r$ in \cref{eq:rhoAB}, we also need to apply measurement operators. It turns out, however, that similar to a teleportation scheme, once unitary corrections, which are assumed error free here, are applied, the output state will not be a function of the measurement outcome. In fact, one can see in \cref{cnot} that for any Bell state at $a_ib_i$ input, the chance of having $|0\rangle$ and $|1\rangle$, at each input is identical. This probability does not change by the unitary operation of CNOT gates, or the identity operator in case of an error, hence right before Z and X-basis measurements on $a_ib_i$, all four possible outcomes are equally likely. Without loss of generality, we then drop the superscript $r$ and calculate the output state for the particular $r$ corresponding to $|0+\rangle_{a_ib_i}$, $i=1,2,3$, for which no Pauli frame correction is needed. We can then apply relevant normalisation factors to \cref{ERC-sum} to
find the joint state $\rho_{AB}$ of memory banks $A$ and $B$, and similarly $C$ and $D$, after remote CNOT operation as follows:
\begin{align}
\label{eq:rhoABsum}
    \rho_{AB} = \frac{1}{2} \sum_{j,k=0,1}\bigotimes_{i=1}^3{\rho_{A_iB_i}^{jk}} ,
\end{align}
where
\begin{equation}
\label{rhoABjk}
    \rho_{A_iB_i}^{jk} = 4{{\rm Tr}_{a_i b_i}[P_0^{a_i}P_+^{b_i} \mathcal{E}_i(|j\rangle_{A_i}\langle k| |0\rangle_{B_i}\langle 0| \rho^{\rm W}_{a_ib_i})]}.
\end{equation}

The next step is to model the ES stage, which can also be thought of certain gate operations, represented collectively by $\mathcal{E}_{\rm ES}$, followed by some single-qubit measurements. In this case, the joint state of memory banks $A$ and $D$, upon observing a measurement outcome $m$ on $B$ and $C$, is given by
\begin{align}
\label{eq:rhoAD}
     \rho_{AD}^m= \frac{U_{AD}^m \text{Tr}_{BC} [M_{BC}^m \mathcal{E}_{\rm ES}(\rho_{\rm ES}^{\rm in}) ] U_{AD}^m}{ p_m},
 \end{align}
where $p_m = \text{Tr} [M_{BC}^m \mathcal{E}(\rho_{\rm ES}^{\rm in}) ]$, $M_{BC}^m$ is the collective projective measurement operator on memory banks $B$, in X basis, and $C$, in Z basis, corresponding to the measurement result $m$, $U_{AD}^m$ is the corresponding Pauli frame correction, and the input state is given by
\begin{equation}
\rho_{\rm ES}^{\rm in} = \rho_{AB} \otimes \rho_{CD},        
\end{equation}
with
\begin{equation}
\label{eq:E2}
    \mathcal{E}_{\rm ES} = \bigotimes_{i=1}^3 \mathcal{E}_{B_iC_i}.
\end{equation}

Using the linear form of the input states as in \cref{eq:rhoABsum}, we then obtain
\begin{equation}
\label{eq:rhoADcal}
    \mathcal{E}_{\rm ES}(\rho_{\rm ES}^{\rm in}) = \frac{1}{4} \sum_{j,k=0,1}\sum_{n,l=0,1} {\bigotimes_{i=1}^3{\mathcal{E}_{B_iC_i}(\rho_{A_iB_i}^{jk}\otimes\rho_{C_iD_i}^{nl})}},
\end{equation}
in which, again, the BSM operation is identical and separable in all rows, and only needs to be calculated once per row in our simulation code. Basically, by breaking the codeword in \cref{linear} into its individual terms, we have broken the entanglement that exists across different rows of \cref{schematic}(b) and can now deal with the state evolution in each row separately. The entanglement will be put together where in the end we add all corresponding terms before applying the decoding operation. 

This whole process, including the imperfect measurement and decoding ones, has analytically been implemented in Mathematica to provide us with an exact description of $\rho_{AD}^{m}$, and its corresponding decoded states, for the first nesting level. The measurement part is straightforward as it also can be implemented horizontally along each row according to \cref{eq:meas}, by which $B$ registers are measured in X basis and $C$ memories are measured in Z basis. That is, in \cref{eq:rhoAD}, we have
\begin{align}
\label{eq:TRBC}
     \text{Tr}_{BC} [&M_{BC}^m \mathcal{E}_{\rm ES}(\rho_{\rm ES}^{\rm in}) ] = \frac{1}{4} \sum_{j,k=0,1}\sum_{n,l=0,1} \bigotimes_{i=1}^3 \rho_{A_iD_i}^{jknl}(m_i), 
   \end{align}
where
\begin{equation}
\label{eq:rhoAiDijknl}
    \rho_{A_iD_i}^{jknl}(m_i) = {{\rm Tr}_{B_iC_i} [ M_{B_iC_i}^{m_i} \mathcal{E}_{B_iC_i}(\rho_{A_iB_i}^{jk}\otimes\rho_{C_iD_i}^{nl})]},
\end{equation}
with $M_{BC}^m = \bigotimes_{i=1}^3 M_{B_iC_i}^{m_i}$ and $m_i$ representing the measurement outcome in row $i$. The decoding process has been implemented by modelling the CNOT gates in the decoding circuit of \cref{decoding} according to \cref{eq:CNOTmodel}. By referring to the whole decoding procedure by operator $\mathcal{E}_{\rm dec}$, we can obtain the final decoded state as follows
\begin{equation}
\label{eq:rho_decode}
    \rho_{m,d}^{\rm dec} = \frac{U_{A_1D_1}^d \text{Tr}_{A_2A_3D_2D_3} [M_{\rm dec}^d \mathcal{E}_{\rm dec}(\rho_{AD}^{m}) ] U_{A_1D_1}^d}{ p_{d|m}},
\end{equation}
where $p_{d|m} =  \text{Tr} [M_{\rm dec}^d \mathcal{E}_{\rm dec}(\rho_{AD}^{m}) ]$, $M_{\rm dec}^d$ is the corresponding measurement operator to outcome $d$ at the decoder ends, and $U_{A_1D_1}^d$ is the corresponding correction operator. 

Computationally speaking, in our method, we are mostly dealing with only 4-qubit systems. This considerably simplifies analytical calculations. There are, however, some exceptions to this. For the 3-qubit repetition code, the last step in \cref{eq:rho_decode} would involve dealing with a 6-qubit system, which is manageable. As the code grows in size, full implementation of the decoding circuit, which requires handling a multipartite entangled state in its input, would become more challenging. In that case, our scheme would still be helpful if we ignore the errors in the decoding circuit. Alternatively, one can think of simpler decoder structures that only rely on single-qubit measurements \cite{AltDecoder}. Imperfect encoding could also cause additional complexity in our technique. In the next subsections, we assess the importance of both encoding and decoding modelling in our analysis. But, before that, let us first explore which measurement outcomes would impact our secret key generation rate the most.

\subsection{Good, bad, and golden states}
\label{good_bad_subsection}

The procedure described above can be used to find the decoded state, $\rho_{m,d}^{\rm dec}$, for any possible outcome $m$ of the ES stage and $d$ of the decoding stage. There are, however, 64 possible values for $m$ and 16 for $d$, each of which could result in a different decoded state, hence different secret key fraction for all those instances that we have got the same measurement outcomes. 

To calculate the total secret fraction, we need to average over all possible outcomes as follows
\begin{align}
\label{eq:r_total}
    r_{\infty}^{\text{total}} = \sum_{m,d} p_{m,d} r_{\infty}^{m,d},
\end{align}
where 
\begin{equation}
p_{m,d} = p_m p_{d|m}  
\end{equation}
is the probability of getting the measurement outcomes $m$ and $d$ and $r_{\infty}^{m,d}= r_{\infty}(\rho_{m,d}^{\rm dec})$ is the secret fraction obtained from \cref{secfraction} and \cref{eq:ezex}. 

\begin{figure}[t]
\includegraphics[width=.8\columnwidth]{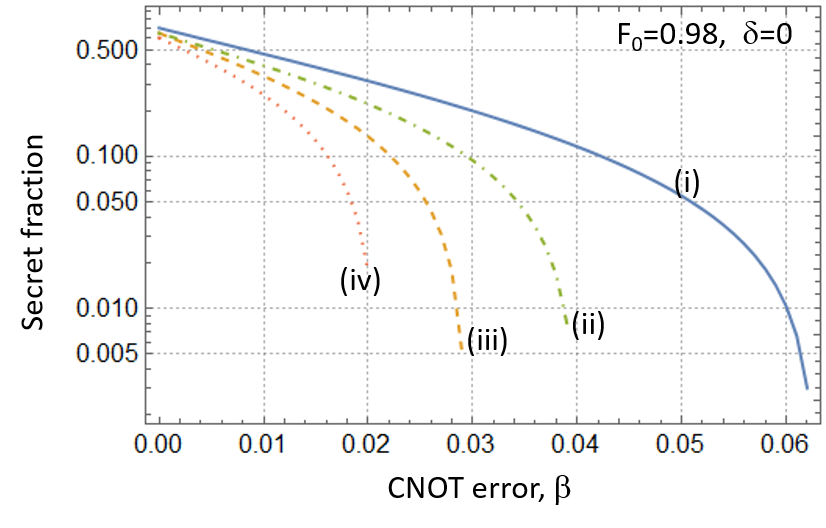}
\caption{\label{Fig:avg} Secret fraction at $F_0=0.98$ and $\delta=0$ for (i) when  we fully use the knowledge of $m$ and $d$, as given by \cref{eq:r_total} (solid blue curve), versus (ii) when only $d$ is known to the users, but not $m$ (dash-dotted green curve), or (iii) when only $m$ is known to the users, but not $d$ (dashed amber curve), or (iv) when none of $m$ and $d$ is used for key extraction (dotted red curve).}
\end{figure}

Note that in \cref{eq:r_total} we make full usage of the available measurement information, $m$ and $d$, from earlier steps. This is expected to give us a higher key rate than the key rate that can be calculated from the state averaged over different ES and/or decoder outcomes. This is because of the convexity of the secret fraction formula in \cref{secfraction} as a function of $e_x$ and $e_z$. Figure~\ref{Fig:avg} confirms this assertion by comparing the secret fraction for the following four cases at $\delta = 0$ and $F_0 = 0.98$ versus $\beta$: (i) when we use the full information in $m$ and $d$ as proposed in this work (solid line); (ii) when we assume the users have no knowledge of $m$, but know $d$, which can be locally obtained by each user (dash-dotted line). In this case, we first find the average ES state over all possible values of $m$, and then pass it to our decoder circuits; (iii) when the users have the information from the ES stage, but the decoder output $d$ will only be used internally in the decoder to correct the shared state (dashed line). In this case, the total secret fraction is given by $\sum_{m} p_{m} r_{\infty}(\rho_{m})$, where $\rho_{m} \equiv \sum_{d}{p_{d|m}\rho_{m,d}^{\rm dec}}$; and (iv) when the users do not know of either $m$ or $d$ before doing QKD measurements (dotted line), i.e., when the decoded state is given by $\rho_{\rm avg} =\sum_{m,d}{p_{m,d}\rho_{m,d}^{\rm dec}}$. In this case, the whole repeater chain and decoders are seen as a black-box channel by the users. As can be seen, by accounting for all different outcomes separately, we can tolerate, respectively, roughly three and two times larger values of $\beta$, as compared to the cases where we use $\rho_{\rm avg}$ or $\rho_{m}$ for secret key extraction. Even if we only use the information at the decoder units, which is at the same place as the users' locations, we can obtain higher key rates than cases (iii) and (iv). This shows the importance of the internal information across the repeater chain and the user boxes in our QKD system. Note that a similar observation has been made for third generation quantum repeaters, and how accounting for syndrome information can boost system performance \cite{Syndrome3gRep}.

The key rate calculation in \cref{eq:r_total} can be cumbersome as many terms need to be considered. There are several ways by which we can group different terms in \cref{eq:r_total} together to reduce the required computation. First, note that, for QKD applications, the secret key analysis is independent of which Bell state is the target state as they are all the same up to local Pauli rotations. Furthermore, the Pauli frame adjustments needed after the BSMs consists of a series of single-qubit operations, which, in our analysis, are assumed perfect. Thus, in this work, we calculate the secret fraction for only $| \tilde{\Phi} \rangle_{A,D}$ as the ES measurement outcome, and use the same result for other encoded Bell states in \cref{eq_es}. This reduces the number of relevant ES outcomes to $16$ corresponding to the measurement results $\{ |+++\rangle_B, |+--\rangle_B, |-+-\rangle_B, |--+\rangle_B \}$, at memory bank $B$, and $\{ |000\rangle_C, |001\rangle_C, |010\rangle_C, |100\rangle_C \}$, at memory bank $C$. Further investigation shows that the four different outcomes at memory bank $B$ do not affect the generated secret fraction as long as the measurement results at memory bank $C$ are the same. We can then only limit ourselves to the specific measurement result $|+++\rangle_B$, which further reduces the number of relevant ES outcomes to $4$. 

Based on the above discussion, we recognise two generic groups of output states, after the ES stage, which we refer to as \textit{good} versus \textit{bad} states. For $| \tilde{\Phi} \rangle_{A,D}$ as the ES measurement outcome, the good ES states correspond to the measurement outcome $|000\rangle_{C}$ where no bit flip has been detected at the ES stage, whereas the bad ES states correspond to the measurement outcomes $|001\rangle_{C}$, $|010\rangle_{C}$, or $|100\rangle_{C}$ in which we have detected a bit-flip error at the ES stage. 

For both good and bad states, we still have 16 cases to consider for the decoder output. We refer to a decoded good state as a  \textit{golden} state if the two users detect no error at their decoder circuits. This corresponds to the measurement outcome $d_g = |00\rangle_{A_2A_3}|00\rangle_{D_2D_3}$. The probability of getting a golden state, and its corresponding total secret fraction is then given by 
\begin{equation}
    p_g = 16 p_{m_g,d_g} {\rm \,\, and \,\,} r_{\infty}^{g}=p_g r_{\infty}^{m_g,d_g},
\end{equation}
where $m_g = |+++\rangle_B|000\rangle_C$, and the factor 16 accounts for the four possible Bell states at the ES stage, and the four outcomes of the $B$ register. Similarly, we have a group of good, but not golden, states, whose corresponding total probability of occurrence and secret fraction are given by
\begin{equation}
     p_{gng} = 16 \sum_{d\neq d_g}{p_{m_g,d}} {\rm \,\, and \,\,}  r_{\infty}^{gng} = 16 \sum_{d\neq d_g}{p_{m_g,d}r_{\infty}^{m_g,d}}.
\end{equation}
Finally the corresponding probability and secret fraction to bad states are given by
\begin{equation}
    p_b = 48 \sum_{d}{p_{m_b,d}} {\rm \,\, and \,\,}  r_{\infty}^{b}= 48 \sum_{d}{p_{m_b,d}r_{\infty}^{m_b,d}} ,
\end{equation}
where $m_b = |+++\rangle_B|100\rangle_C$, and the factor 48 covers three different locations of a single error in register $C$, each at 16 different cases of Bell state and $B$ register outcomes as in golden states. The total secret fraction is then given by
\begin{equation}
\label{eq:rinftcomp}
 r_{\infty}^{\rm total}= r_{\infty}^{g}+ r_{\infty}^{gng} + r_{\infty}^{b}.
\end{equation}

One of the key results of this work is to show that, in most practical cases, the golden states are the main positive contributor to the key rate formula in \cref{eq:rinftcomp}, that is, $r_{\infty}^{\text{total}} \gtrapprox r_{\infty}^{g}$. This result allows us to considerably reduce the complexity of the problem in that, instead of accounting for all possible outcomes at different parts of the repeater chain, we only focus on a single class of states.  

\begin{figure}[t]
\includegraphics[width=\columnwidth]{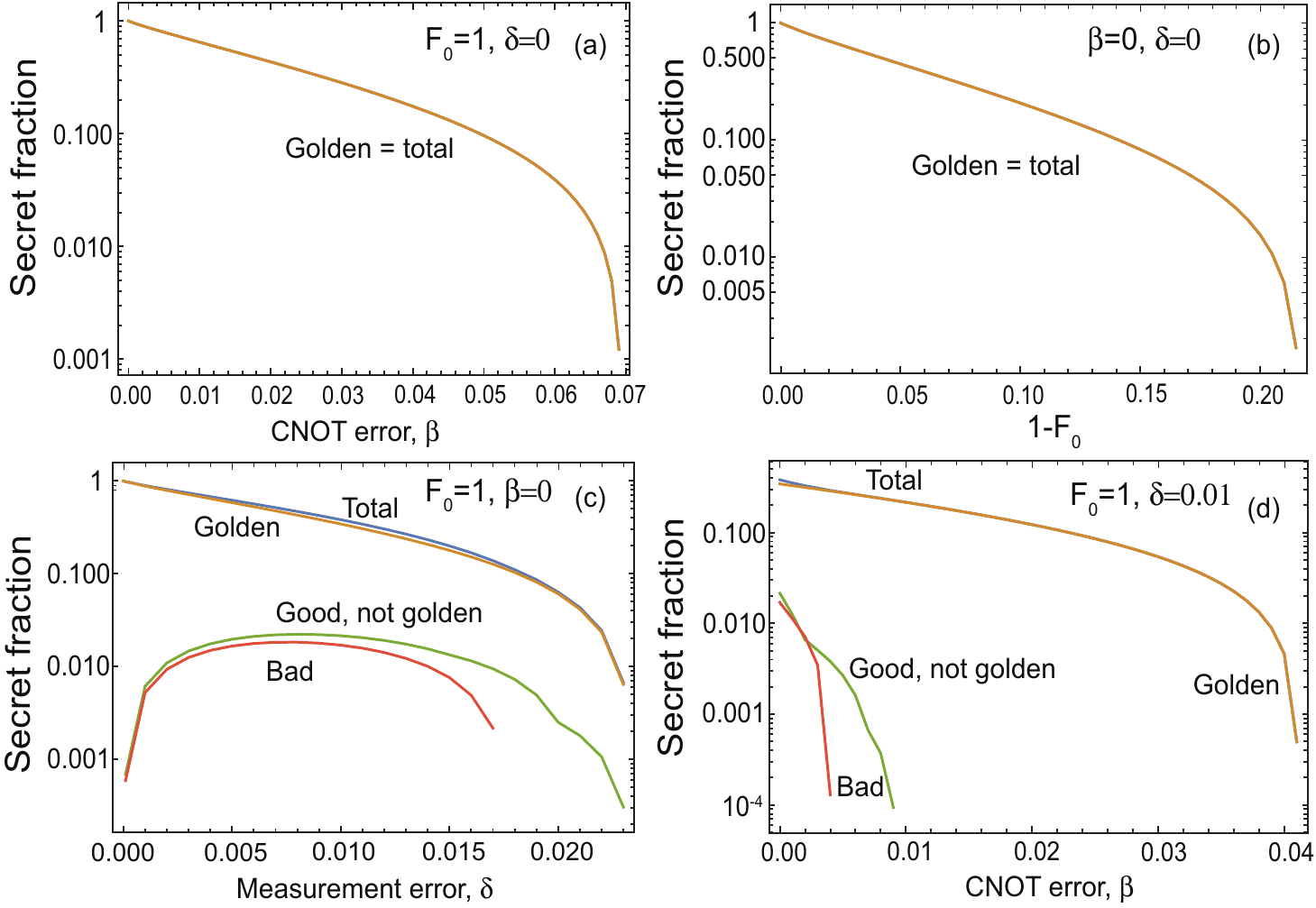}
\caption{\label{good_bad} Secret fraction as a function of different error parameters at (a) $F_0=1$ and $\delta=0$; (b) $\beta =0$ and  $\delta=0$; (c) $F_0=1$ and $\beta =0$; and (d) $F_0=1$ and  $\delta=0.01$. In all graphs, the top blue curve is for the total secret fraction, $r_{\infty}^{\rm total}$, followed by the golden curve, $r_{\infty}^{g}$, for golden states, the green curve $r_{\infty}^{gng}$ for good, but not golden, states, and the red curve, $r_{\infty}^{b}$, for bad states. In (a) and (b), the latter two terms are zero, so the golden curve overlaps with the blue one.}
\end{figure}

Here, we demonstrate how different kinds of states contribute to the key rate. Figure~\ref{good_bad} shows the total secret fraction and its three main components in \cref{eq:rinftcomp} for different parameter regimes, for the initial codeword states as in \cref{linear}. We make several interesting observations from this figure, as summarized below:
\begin{itemize}
    \item {\bf Observation 1:} At $\delta=0$, only golden states can generate positive key rates. This has been shown in Figs.~\ref{good_bad}(a)-(b). In Fig.~\ref{good_bad}(a), we have assumed that the initial Bell states are ideal and that there is no measurement error. We have then plotted the secret fraction versus the CNOT gate error parameter, $\beta$. It can be seen that, in this case, the golden state is the only contributor to the total secret fraction. It turns out that for all other states the phase error rate is at its worst possible value of 0.5 at which no secret key can be generated. We have a similar observation in Fig.~\ref{good_bad}(b), where, now, $\beta = 0$, and $F_0$ is a variable. In this case, our analysis indicates that most decoder outcomes simply never happen. But, even if they do, except for golden states, for all other terms $e_x =0.5$. To see why this happens we can look back at the ideal state obtained after the ES operation in \cref{eq_es}. In order to detect an error state such as $|+++\rangle_B|100\rangle_C$ at the ES stage, we can either have an error corresponding to $X_{C_1}$, which results in $|\phi^+\rangle$ after decoding, or something like $Z_{B_1}X_{C_1}$, which results in $|\phi^-\rangle$. If we trace back these errors, using known circuits that convert an error after a CNOT gate to errors before it \cite{bratzik2014}, we can see that such errors, respectively, originate from $|\psi^+\rangle$ and $|\psi^-\rangle$ somewhere earlier in the circuit. In the case of imperfect Bell states, this is caused by the terms in the input Werner state in \cref{original_bell}. The identity operator in the imperfect CNOT gate can similarly introduce such states in the circuit resulting in a similar behavior. In both cases, the weight of $|\psi^+\rangle$ and $|\psi^-\rangle$ is the same at the input mixture, resulting in an equal mixture of $|\phi^+\rangle$ and $|\phi^-\rangle$ after decoding. At $\delta =0$, according to \cref{eq:ezex}, this results in $e_x =0.5$.  
    \item {\bf Observation 2:} At $\delta \neq 0$, non-golden states can contribute to the total secret fraction but at comparatively much lower values. This can be seen from Figs.~\ref{good_bad}(c)-(d). In Fig.~\ref{good_bad}(c), we have fixed $\beta$ and $F_0$ to their ideal values and have plotted the secret fraction for different values of $\delta$. This is the first case in which $r_{\infty}^{gng}$ and $r_{\infty}^{b}$ take nonzero values for some values of $\delta$. The reason for this is that if we detect an error at the ES stage, for instance, by observing $|100\rangle_C$, because of measurement errors, the actual state in hand, according to \cref{eq_es}, is most likely still the ideal state. Most cases for the decoder output are also similarly benign. The errors that may happen at the remote CNOT stage could equally result in bit or phase-flip errors, both with a probability scaling with $\delta$. This allows us to have positive key rates for bad, as well as, good, but not golden, states. At low values of $\delta$, however, the overall chance of obtaining such states is much lower than that of the golden states, which makes the total secret fraction still approximately the same as $r_{\infty}^g$. Finally, in Fig.~\ref{good_bad}(d), we have verified this finding when $\beta$ and $\delta$ are nonzero. We have chosen $\delta = 0.01$ as it gives a high rate for bad states in \cref{good_bad}(c). We observe that the key rate for bad and good-but-not-golden states is nonzero for small value of $\beta$. This suggests that so long as the phase error rate is dominated by the measurement error we can get a positive key rate for non-golden states. But, once $\beta$ increases to the level that the dominant source of phase error is what we discussed in Observation 1, then no secret keys can be extracted from such terms. At $\delta = 0.01$ the onset of dominance of CNOT errors is just before $\beta=0.01$. At $\delta = 0.001$, we have verified that the golden state is the only contributor to the key rate for $\beta>0.0046$. Given that in practice it is easier to have a low value for $\delta$ as compared to $\beta$, this observation suggests that for sufficiently small $\delta$, the errors in the two-qubit gates would be the dominant factor in determining the final key rate. The latter can reliably be calculated from golden states in such cases.
    \item {\bf Observation 3:} At cut-off point, the golden states are the main contributor to the key rate. Even though the non-golden states can contribute to a small extent to the key rate within some range of parameters, their contribution effectively ceases to zero by the time that we get to the cut-off point for our QKD system. This suggests that to find such maximum allowed error rates, one can reliably only calculate the key rate for the golden states.
    \item {\bf Observation 4:} The measurement error $\delta$ has the lowest cut-off point ahead of $\beta$ and $1-F_0$. According to \cref{good_bad}(c), at $\delta \approx 0.023$, the key rate drops to zero. This happens at $\beta \approx 0.07$, in \cref{good_bad}(a), and $F_0 \approx 0.76$ in \cref{good_bad}(b). This could simply be because of the number of measurement operations in the whole setup exceeding the number of CNOT gates. But, this also suggests that unless $\delta$ is sufficiently small, its effect cannot necessarily be neglected in a reliable analysis of the system. 
\end{itemize}

Based on the above observations, in the remaining of this paper, we only calculate $r_{\infty}^{g}$. This is a tight lower bound on $r_{\infty}^{\rm total}$, in line with the common practice in calculating the key rate in QKD. More importantly, this suggests a practical distillation technique in such encoded repeaters, in which one can simply ignore the output if any error has been detected at the ES or decoding stage. This could substantially simplify the implementation of such systems in their early demonstrations. Under the assumption that $r_{\infty}^{g}$ closely follows $r_{\infty}^{\rm total}$, this distillation technique is more effective than relying on the error correction capabilities of the code. That is, in practical QKD settings, we may only need to use the error detection features of a code rather than its error correction power.

\subsection{The effect of the encoding and decoding circuits on the secret fraction}
\label{encoding_decoding_subsection}

In this section, we study how errors in encoding and decoding circuits would affect the achievable secret fraction. Thus far, we have only considered the perfectly encoded states as given by \cref{linear}, which can be a reasonable assumption if one uses probabilistic techniques to initialize the memories. If, however, one uses CNOT gates to create such states deterministically, we should also account for errors in such gates.  In this case, the initial codeword states for memory bank $A$, as an example, is given by \cite{bratzik2014}
\begin{equation}
\label{rhofull}
    \rho_A^{\rm in} = \rho_A^{\rm code} + \rho_A^{\rm other},  
\end{equation}
where
\begin{align}
\rho_A^{\rm code} &= \frac{1}{2} [ 1+\beta(\beta/2-5/4) ] (|000\rangle_A \langle 000| +  |111\rangle_A \langle 111| ) \nonumber\\
              &\, +\frac{1}{2} (1-\beta)^2 (|000\rangle_A \langle 111| +  |111\rangle_A \langle 000| )  \label{imperfect_encoding_part} 
\end{align}
and
\begin{align}
\rho_A^{\rm other} &= \frac{\beta}{4}(3/2-\beta) (|101\rangle_A \langle 101| +  |010\rangle_A \langle 010| ) \nonumber \\
              &\, +\frac{\beta}{8} (|001\rangle_A \langle 001| + |100\rangle_A \langle 100| ) \nonumber \\
              &\, +\frac{\beta}{8} ( |110\rangle_A \langle 110| +  |011\rangle_A \langle 011| ).  \label{imperfect_encoding_all}
\end{align}
The terms in \cref{imperfect_encoding_part} are effectively the encoded state in \cref{linear} although with modified weights to account for CNOT errors. Our linearization technique is easily applicable to these terms as they are still in the desired tensor product form of having the same input qubit in all rows. To apply our technique to the other terms in \cref{imperfect_encoding_all}, we need to consider many more combinations of input states, which will increase the complexity of the simulation especially at higher nesting levels. Here, through the comparison of the secret fraction for different input states, we show that the coded part in \cref{imperfect_encoding_part} plays the major role in determining the secret fraction, based on which we can neglect the other terms. This will crucially simplify the code for further simulation. Note that the above states have been obtained by first applying $\mathcal{E}_{A_1A_2}$ and then $\mathcal{E}_{A_1A_3}$, as the two operators do not commute for nonzero values of $\beta$.

\begin{figure}[t]
\includegraphics[width=0.8\columnwidth]{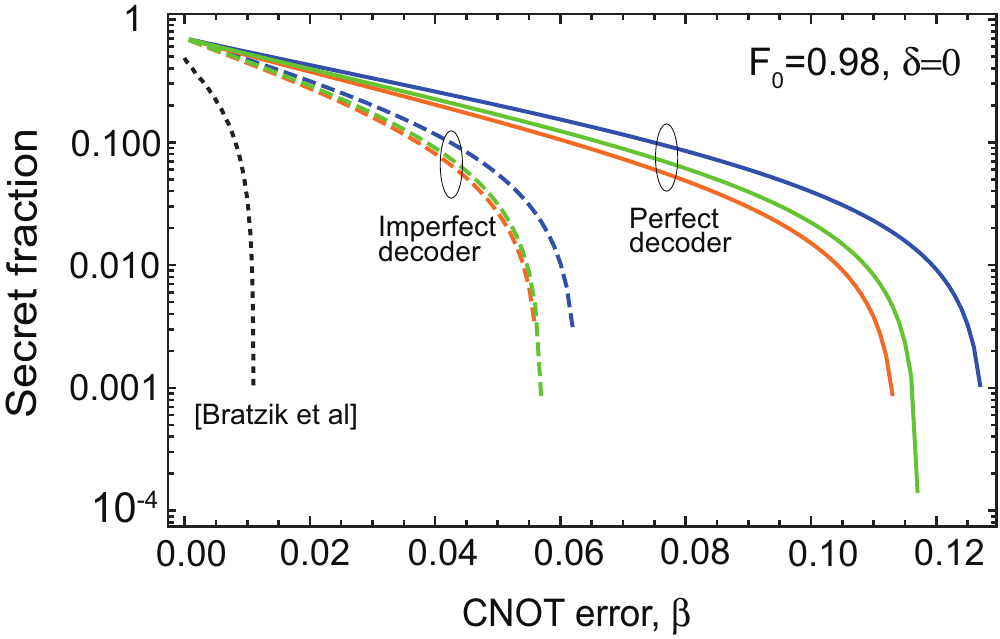}
\caption{\label{1st_nesting_level}  Secret fraction versus $\beta$ at $F_0=0.98$ and $\delta=0$. The solid lines correspond to error-free decoding circuits, and the dashed lines correspond to imperfect decoding circuits. The top blue curve in each batch corresponds to ideal encoding; the lower orange lines correspond to imperfect encoders as modelled by \cref{rhofull}, and the middle green lines correspond to the coded part of the encoded state given by \cref{imperfect_encoding_part}. In all cases the secret fraction is lower bounded by that of the golden states. The black dotted curve is the corresponding graph obtained in \cite{bratzik2014} for the same parameter values for their model of imperfect encoders and decoders. }
\end{figure}

Figure~\ref{1st_nesting_level} shows the secret fraction versus $\beta$ at $F_0 = 0.98$ and $\delta =0$ in several different cases. The top three curves (solid lines) give the secret fraction if we neglect all sources of error at the decoder stage, whereas the next batch of three curves (dashed lines) account for errors in the decoder circuit. In each batch, we consider three cases: (i) the encoding circuits are all perfect (top blue curves), that is, we assume $\beta =0$ in these modules; ({ii}) The encoding process is modelled by the imperfect encoded state given in \cref{rhofull} (the bottom orange curves); and ({iii}) The encoding process is modelled by the state $\rho_A^{\rm code}$ in \cref{imperfect_encoding_part} (green curves in the middle of the circled batches). Two important observations can be made from this graph. First, it is clear that the imperfections in the decoder module is far more important than the encoder one. For a realistic analysis of the system, it will then be crucial to account for decoder errors, as we do in this paper. The second point is that, especially in the case of imperfect decoders, which is of practical interest, the effect of $\rho_A^{\rm other}$ on the secret fraction is effectively negligible, as the curve obtained from $\rho_A^{\rm code}$ very closely follows that of the imperfect encoder modelled by the full state in \cref{rhofull}. In the rest of this work, we will then only account for $\rho_A^{\rm code}$ when we model imperfections in the encoders. As mentioned earlier, this will substantially simplify our analysis as we only need to replace $\rho_{A_iB_i}^{jk}$ in \cref{rhoABjk} with $(C_{jk})^{1/3} \rho_{A_iB_i}^{jk}$, where $C_{00} = C_{11} =  1+\beta(\beta/2-5/4)$ and $C_{01} = C_{10} =  (1-\beta)^2$.

In \cref{1st_nesting_level}, we have also compared our results with Fig.~6 in \cite{bratzik2014}, which, for the same parameters, obtains the secret fraction for the same system but without using the post-selection that we make on the basis of good/bad states, or decoder outputs. The corresponding curve in \cite{bratzik2014} is shown by the dotted black line. The results clearly demonstrate how substantially one can improve the performance of QKD over encoded repeater setups by relying mainly on the error detection, rather than correction, features of the code. This could also change the main conclusion drawn in \cite{bratzik2014} in that such repeaters can hardly outperform other classes of deterministic repeaters as the cut-off point for $\beta$ has nearly improved by six folds from nearly 0.01 to about 0.06 when imperfections in both encoders and decoders are considered. Another distinction, between our work and that of \cite{bratzik2014} is in the way errors have been modelled in each case. In \cite{bratzik2014}, errors are modelled collectively by an identity operator even if there is a cascade of operations. This is expected to overestimate the error in the system. In our work, we account for errors per individual gates, which gives us a more accurate picture of how errors propagate to the final state, and eventually affect the secret fraction. Based on the findings in \cref{Fig:avg} and \cref{1st_nesting_level}, there is a two-fold improvement in the cut-off value of $\beta$ because of such more accurate modelling and calculations.

In the following section, we use the results of this section to analyse the repeater chain at higher nesting levels. Based on the performance analysis for nesting level one, we will only consider the golden state contribution to the secret fraction. Unless otherwise mentioned, we fully account for imperfections in the decoder, but only use the coded components in \cref{imperfect_encoding_part} to model the encoder.

\section{Extension to higher-nesting levels}
\label{higher_nesting_level_section}
The methodology developed in \cref{linearization_section} can be extended to higher nesting levels in a recursive way. For instance, at nesting level $n=2$, we can think of 8 memory banks named $A$ to $H$, where we first apply our ES technique to $BC$ and $FG$ pairs and then $DE$. In this case, the output state of the ES stage, for measurement output $m_g = |+0\rangle$ at all corresponding ES measurements, can be written as follows:
\begin{equation}
    \rho_{AH}^{m_g} =\rho_{\rm ES}^{(2)}/{\rm Tr}[\rho_{\rm ES}^{(2)}],
\end{equation}
where
\begin{equation}
    \rho_{\rm ES}^{(2)} = \frac{1}{16} \sum_{j_1,\ldots, j_8=0,1} {\bigotimes_{i=1}^3 \rho_{A_iH_i}^{j_1,\ldots, j_8}(m_g)},
\end{equation}
with 
\begin{align}
    \rho_{A_iH_i}^{j_1,\ldots, j_8}(m_g) &= {\rm Tr}_{D_iE_i} [ M_{D_iE_i}^{m_g} \nonumber \\
    &\mathcal{E}_{D_iE_i}(\rho_{A_iD_i}^{j_1,\ldots, j_4}(m_g) \otimes \rho_{E_iH_i}^{j_5,\ldots, j_8}(m_g))].
\end{align}
Here, $\rho_{A_iD_i}^{j_1,\ldots, j_4}(m_g)$ and $\rho_{E_iH_i}^{j_5,\ldots, j_8}(m_g)$ have already been calculated in \cref{eq:rhoAiDijknl}. One can generalize this technique to higher nesting levels in a similar way to obtain the corresponding matrix $\rho_{\rm ES}^{(n)}$ for nesting level $n$. The corresponding golden state for the two end nodes $A$ and $A'$ is then given by 
\begin{equation}
    \rho_{AA'}^{(n)} = \rho^{(n)}_{\rm dec}/{\rm Tr}[\rho^{(n)}_{\rm dec}],
\end{equation}
where
\begin{equation}
    \label{eq:rho_decode_n}
    \rho^{(n)}_{\rm dec} = \frac{\text{Tr}_{A_2A_3A'_2A'_3} [P_0^{A_2} P_0^{A_3} P_0^{A'_2} P_0^{A'_3} \mathcal{E}_{\rm dec}(\rho_{\rm ES}^{(n)}) ]} { {\rm Tr}[\rho_{\rm ES}^{(n)}]}.
\end{equation}
The corresponding secret fraction can then be lower bounded by
\begin{equation}
    r_\infty^{(n)} = 16^{2^n-1} {\rm Tr}[\rho_{\rm ES}^{(n)}] {\rm Tr}[\rho^{(n)}_{\rm dec}] r_\infty( \rho_{AA'}^{(n)}),
\end{equation}
where the prefactors are, respectively, the number of golden states at nesting level $n$ and the corresponding probability for each.

\begin{figure}[t]
\includegraphics[width=\columnwidth]{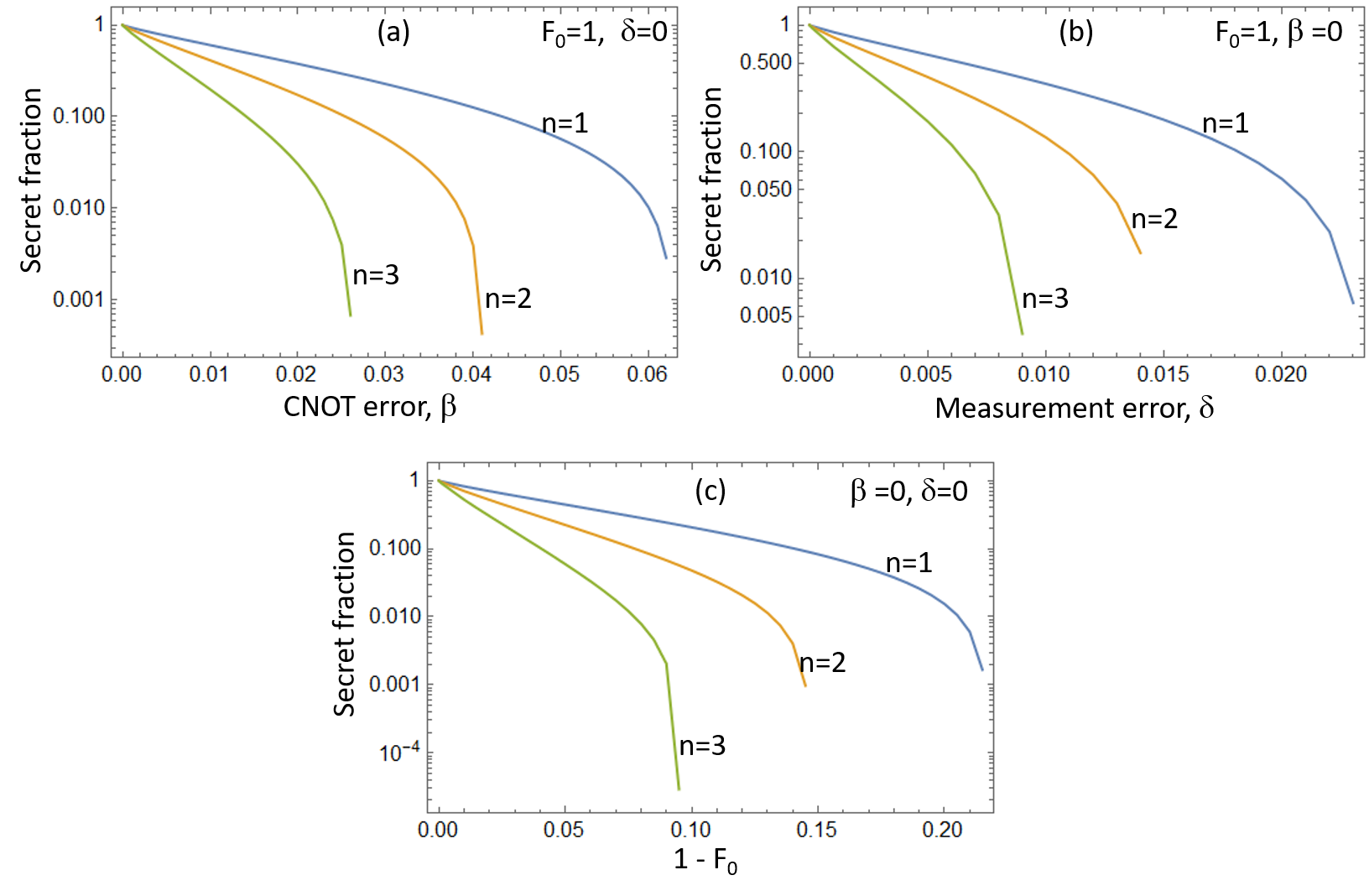}
\caption{\label{nesting_level1-3} The secret fraction as a function of (a) gate error probability $\beta$, (b) measurement error probability $\delta$, and (c) the error in the initial Bell states $1-F_0$, at different nesting levels. In each case, the other two parameters have taken their ideal values.}
\end{figure}

As an application of the analytical method we developed above, we look into the dependence of the secret fraction on various sources of errors in the setup. Figure ~\ref{nesting_level1-3} shows the secret fraction, for the first three nesting levels, as a function of $\beta$, $\delta$, and $1-F_0$, while, in each case, the other two parameters are assumed ideal. As expected, the secret fraction drops as we go to higher nesting levels as the number of gates and measurement operations exponentially grows with the nesting level. The resilience to error parameters would correspondingly go down, but, instead, we are covering exponentially longer distances, at higher nesting levels, if we assume the elementary link is of the same length in all cases. Given that by increasing the nesting level by one, we have over twice as many operations as before, a simple rule of thumb may suggest that the cut-off point for each source of error must be halved. Our exact calculations suggest that the new cut-off points are slightly better than what is predicted by this rule of thumb, which could be because some errors cancel each other when one considers all possibilities, as we do in our analysis. For instance, at $n=1,2,3$, the maximum allowed $\beta$ is, respectively, 0.062, 0.041, and 0.026. As it was the case for $n=1$, the secret fraction is most sensitive to $\delta$ and least sensitive to $F_0$. It is therefore crucial to have accurate single-qubit measurement operations in such QRs to make them useful for QKD purposes.

\section{Conclusions and Discussion}
\label{conclusion_section}
In this work, we studied the performance of QKD systems run over a repeater setup that used three-qubit repetition codes for entanglement distillation. By modeling the error in all two-qubit gates and single-qubit measurements, we obtained an accurate picture of the requirements of such systems. It turned out that such systems could considerably be more resilient to errors than previously thought. The system was most sensitive to measurement errors, but, provided that they were kept sufficiently low in the experimental setup, we showed that CNOT errors on the order of a few percents could be tolerated. The QKD system could also handle imperfections in the initial Bell states aligned with what experimentally is achievable today \cite{dolde2014high,casabone2013heralded}. To handle the computational complexity associated with this many-qubit repeater setup, we devised an analytical technique for modelling the repeater chain, where, at the core of it, we only needed to deal with four qubits at a time. This enabled us to obtain the analytical form of the final entangled states shared between the two end users after several nesting levels. Moreover, our analysis enabled us to fully account for the information available to the end users, from entanglement swapping and decoding circuits, in their secret key distillation. By using this information, we showed three-fold increase in resilience to errors in CNOT gates as compared to when the repeater chain and decoders are treated as a black box. By looking at different sets of measurement outcomes, we then identified the key \textit{golden} states that contributed the most to the final key rate. These golden states corresponded to the cases where no error had been detected at entanglement swapping and decoding stages. This observation resulted in a simple, but effective, post-selection tool for our QKD system that entirely relied on the error detection features of the code, rather than its error correction as when we treat the repeater chain as a black box. We also studied the impact of errors in the encoder and decoder circuits and showed that the latter is much more detrimental to the QKD system. 

The analytical framework derived in this paper can be improved and extended to consider more complex code structures and alternative decoders. One of the computational challenges that we have to deal with is the number of terms that needs to be calculated in the final state. In its exact form, we need to consider all combinations of input states to the elementary links, whose number grows exponentially with the nesting level. To manage the complexity, we need then to identify which input combinations have a major impact on the final key rate, and which ones could perhaps be neglected for a tight approximation. The decoder setup could also pose computational challenges as in its current form, it takes a multi-qubit entangled state at its input and gives a bipartite state at its output. For large codes, it may be hard to computationally handle the large input. Alternative decoders may need to be designed to offer competitive performance especially if larger codes suffer more from errors in the system. Finally, this work mainly relied on finding the key rate once the repeater chain had generated an entangled state. In order to calculate the total key rate one should look at the timing of the protocol with respect to the initial entanglement distribution and how multiplexing is used in the system. All the above will be addressed in a forthcoming paper under preparation \cite{AltDecoder}.  

\begin{acknowledgments}
The authors would like to thank Dagmar Bru{\ss}  and Hermann Kampermann for fruitful discussions. This project is funded by the European Union's Horizon 2020 research and innovation programme under the Marie Sklodowska-Curie grant agreement number 675662 (QCALL) and UK EPSRC Grant EP/M013472/1. All data generated in this work can be reproduced by the provided methodology and equations.
\end{acknowledgments}


\end{document}